\documentclass[prl,twocolumn,longbibliography]{revtex4}

\usepackage[utf8]{inputenc}
\usepackage{bm}
\usepackage{mathtools}
\usepackage{bbold}
\usepackage{amsmath}
\usepackage{amssymb}
\usepackage{hyperref}
\usepackage[normalem]{ulem}
\usepackage{color}
\usepackage[dvipsnames]{xcolor}

\begin{document}

\title{Entangling dynamics from effective rotor/spin-wave separation \\
in U(1)-symmetric quantum spin models}
\author{Tommaso Roscilde, Tommaso Comparin, Fabio Mezzacapo}
\affiliation{Univ Lyon, Ens de Lyon, CNRS, Laboratoire de Physique, F-69342 Lyon, France}


\begin{abstract}
The non-equilibrium dynamics of quantum spin models is a most challenging topic, due to the exponentiality of Hilbert space; and it is central to the understanding of the many-body entangled states that can be generated by state-of-the-art quantum simulators. A particularly important class of evolutions is the one governed by U(1) symmetric Hamiltonians, initialized in a state which breaks the U(1) symmetry -- the paradigmatic example being the evolution of the so-called one-axis-twisting (OAT) model, featuring infinite-range interactions between spins. In this work we show that the dynamics of the OAT model can be closely reproduced by systems with power-law-decaying interactions, thanks to an effective separation between the zero-momentum degrees of freedom, associated with the so-called Anderson tower of states, and reconstructing a OAT model; and finite-momentum ones, associated with spin-wave excitations. This mechanism explains quantitatively the recent numerical observation of spin squeezing  and Schr\"odinger-cat generation in the dynamics of dipolar Hamiltonians; and it paves the way for the extension of this observation to a much larger class of models of immediate relevance for quantum simulations.  
\end{abstract}
\maketitle

\emph{Introduction.}
The controlled generation of many-body entangled states \cite{Horodeckietal2009,Guehne_2009,Pezze2018RMP,Friisetal2019} is the central feature of quantum many-body devices, such as quantum simulators \cite{Georgescuetal2014}, quantum computers \cite{IkeandMike} and entanglement-assisted sensors \cite{Pezze2018RMP}, based \emph{e.g.} on neutral atoms \cite{Gross2017,BrowaeysL2020,Schaeferetal2020,Kaufman2021NP}, trapped ions \cite{Monroeetal2021} or superconducting circuits \cite{Juanjobook}. 
Identifying realistic and robust protocols for the scalable preparation of multipartite entangled states \cite{Pezze2018RMP} is essential for fundamental studies on quantum matter using such devices, as well as for their most advanced applications. 
In this work we shall specialize to the paradigm of analog quantum simulation based on time-independent Hamiltonians ${\cal H}$, which generically takes as an input a fiducial, non-entangled initial state $|\psi(0)\rangle$, and transforms it into an entangled state $|\psi(t)\rangle = \exp(-i{\cal H} t)|\psi(0)\rangle$ via the global unitary evolution. Among the numerous many-body Hamiltonians realizable with state-of-the-art simulators, it is crucial to identify those giving rise to entanglement which can be produced and certified with \emph{polynomial resources} -- namely produced after evolution times $t$ scaling polynomially with system size, and certified via standard observables requiring a polynomial amount of statistics. A further desideratum is for entanglement to be \emph{scalable}, namely multipartite and with a depth scaling with system size, offering in this way a fundamental test of the scalability of quantum superpositions; as well as the central resource  for \emph{e.g.} entanglement-assisted metrology \cite{Pezze2018RMP}. These properties are far from being trivial: generic many-body Hamiltonians evolving random initial factorized states lead to extensive entanglement entropies, which nonetheless can only be certified using exponentially scaling resources \cite{Kaufmanetal2016,Brydges2019}. The above requirements can instead be met by exploiting special symmetry properties (exact or approximate) of the many-body Hamiltonian.  

 A paradigmatic example of an entangling many-body Hamiltonian giving rise to multipartite entangled states with scalable depth is offered by the one-axis-twisting (OAT) model \cite{Kitagawa1993PRA}. The latter describes an ensemble of spins of length $S$ interacting via infinite-range interactions leading to planar-rotor Hamiltonian, ${\cal H}_{\rm OAT} = (J^z)^2/(2I)$. Here $J^{\alpha} = \sum_{i=1}^N S_i^\alpha$ ($\alpha=x,y,z$) is the collective-spin operator, and $I\sim N$ is the macroscopic moment of inertia of the rotor. When the dynamics is initialized in the coherent spin state $|{\rm CSS}_x\rangle = |\psi(0)\rangle = \otimes_{i=1}^N |\rightarrow_x\rangle_i$ polarized along the $x$ axis, the evolved state $|\psi(t)\rangle$ develops first spin squeezing, characterized by a squeezing parameter \cite{Wineland1994PRA}  $\xi_R^2 = N \min_{\perp}{\rm Var}(J^\perp)/\langle J^x \rangle^2$ -- where $\min_{\perp}$ indicates the minimization over the collective-spin components in the $yz$ plane, perpendicular to the average spin orientation. A parameter $\xi_R^2 < 1/k$ witnesses $(k+1)$-partite entanglement. During the OAT dynamics it reaches a minimal value $\left ( \xi_R^2 \right)_{\rm min} \sim N^{-\nu}$ (with $\nu = 2/3$ for very large system sizes)   at a time $t_{\rm sq} \sim N^{1/3}$, realizing therefore scalable multipartite entanglement \cite{Kitagawa1993PRA,Wineland1994PRA}. Moreover, at times $t_q = 2\pi I/q$  (for $N$ even, and $q=2, ..., q_{\rm max}$ with $q_{\rm max} \sim \sqrt{N}$) the OAT dynamics realizes a cascade of $q$-headed Schr\"odinger's cat states \cite{Agarwal1997PRA}, namely superpositions of $q$ CSS states rotated around the $z$ axis by integer multiples of $2\pi/q$ with respect to the initial $|{\rm CSS}_x\rangle$. This sequence culminates with the $q=2$ (or GHZ) state $(|{\rm CSS}_x\rangle + i |{\rm CSS}_{-x}\rangle)/\sqrt{2}$ at time $\pi I$, featuring $N$-partite entanglement.

  \begin{figure*}[ht!]
\includegraphics[width=0.9\textwidth]{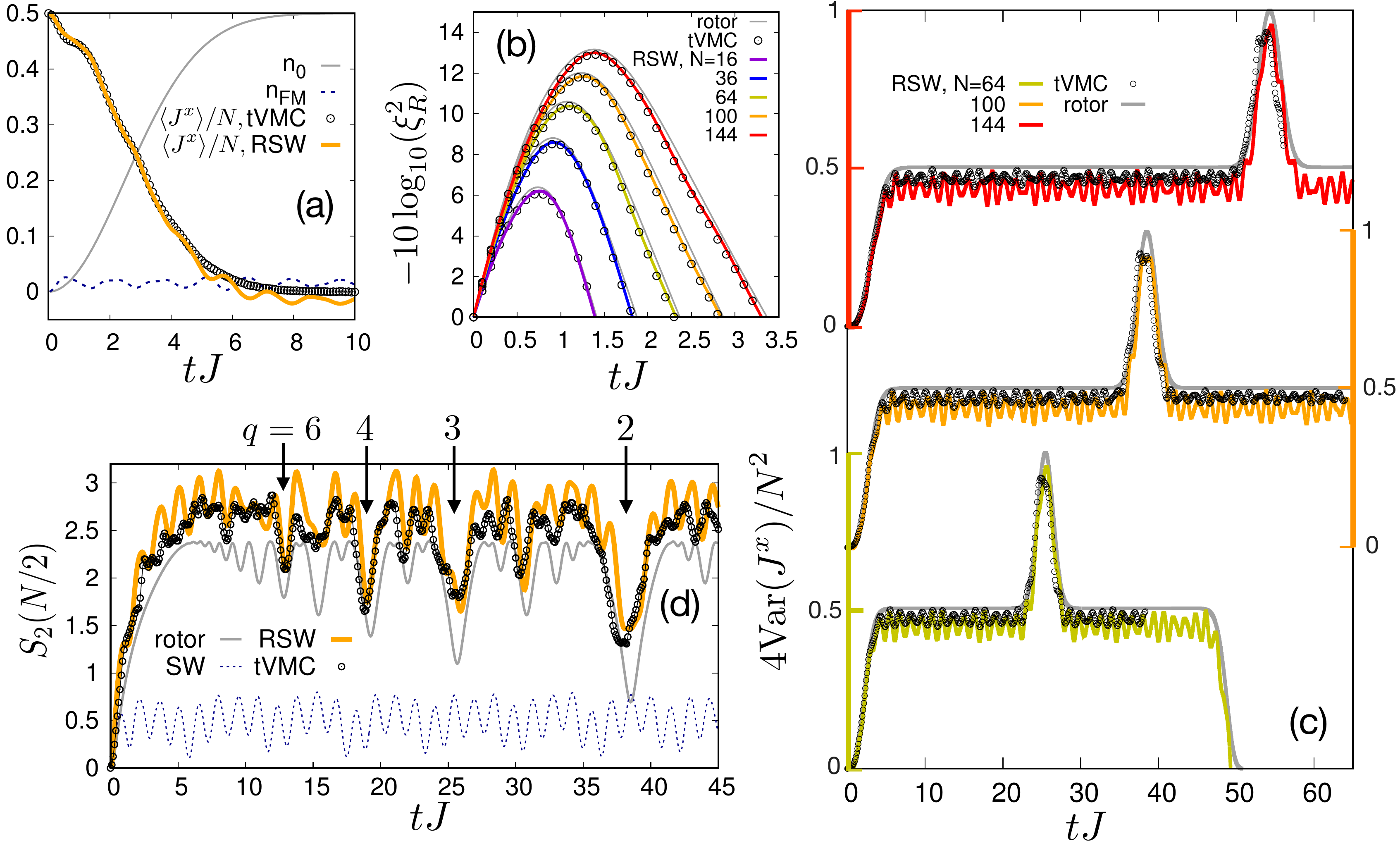}
\caption{\emph{Dynamics of the 2$d$ dipolar XX model.} (a) Dynamics of the average magnetization $\langle J^x \rangle$ for $N=100$ spins, comparing RSW and tVMC ones. The graph shows as well the density of zero-momentum bosons, $n_0 = \langle b_0^\dagger b_0 \rangle/N$, and of finite-momentum ones $n_{\rm FM} = N_{\rm FM}/N$; (b) Spin squeezing parameter for various system sizes; in this and further panels, the rotor results correspond to those of a OAT model with moment of inertia $I_{\rm ToS}$. (c) Dynamics of the magnetization variance; (d) Dynamics of the half-system R\'enyi entropy for a system of $N=100$ spins, showing the separate rotor and SW contributions whose sum leads to the RSW prediction. The arrows mark the time of appearance of some of the $q$-headed cat states.}
\label{fig1}
\end{figure*}
 
 The OAT model with its infinite-range interactions can be realized literally using atoms or superconducting qubits \cite{Riedel2010,Gross2012,Norciaetal2018,Bravermanetal2019,Song2019}; and the full sequence of the above-cited entangled states has been realized in recent experiments \cite{Song2019}. Nonetheless, there is mounting evidence that the same dynamics can be obtained using a wide variety of models, such as systems of qubits ($S=1/2$ spins), provided that their interactions are sufficiently long-ranged, and share with the OAT model its fundamental U(1) symmetry. 
 OAT-like squeezing dynamics has been theoretically reported in XXZ models with power-law-decaying interactions  \cite{FossFeig2016Arxiv,Perlin2020PRL,Comparin2022PRA,Comparinetal2022b,Blocketal2023}; and the formation of the whole cascade of $q$-headed cat states has been demonstrated by us for dipolar interactions in 2$d$ \cite{Comparinetal2022b}. Yet a quantitative understanding of the persistence of OAT-like dynamics beyond the OAT model is still lacking, in spite of its fundamental importance in order to establish many-body Hamiltonians as potential resources of scalable multipartite entanglement. 
 
 In this work we offer a quantitative theoretical insight into this problem, by highlighting an effective mechanism of \emph{separation of variables} taking place in a broad class of models with U(1) symmetry. Making use of a spin-boson mapping, the spin degrees of freedom can be decomposed into a zero-momentum component, reconstructing an effective OAT model when all non-linearities are properly accounted for; and finite-momentum components, which reconstruct linear spin-wave (SW) excitations at the lowest order in the expansion of the Hamiltonian in powers of bosonic operators. Neglecting the coupling between zero-momentum and finite-momentum bosons (justified when SW excitations are weakly populated) leads to a rotor/spin-wave (RSW) separation scheme:  this scheme predicts an additive structure of most salient observables and entanglement entropies, justifying how the full-fledged OAT dynamics can emerge in systems with spatially decaying interactions. The predictions of RSW theory are quantitatively confirmed by time-dependent variational Monte-Carlo (tVMC) results in the relevant case of dipolar interactions in two-dimensions -- for which tVMC is extremely as shown by us in Ref.~\cite{Comparinetal2022b}.  
 
 \emph{From spins to bosons; rotor/spin-wave separation.} In this work we focus on the XXZ model for quantum spin lattices
 \begin{equation}
{\cal H}_{\rm XXZ} = - \sum_{i< j}  J_{ij} \left ( S_i^x S_j^x + S_i^y S_j^y + \Delta S_i^z S_j^z  \right) 
\label{e.XXZ}
\end{equation}
where $J_{ij}$ is an arbitrary matrix of ferromagnetic couplings, $J_{ij} \geq 0$; and $\Delta$ is the anisotropy parameter. Throughout the rest of this work the sites $i,j$ are defined on a periodic lattice with $N=L^d$ sites in $d$ dimensions.  In order to quantitatively relate the XXZ model to the OAT one, we first map locally the spins onto Holstein-Primakoff (HP) bosons, $S_i^x =  S - n_i$, $S^y_i = (\sqrt{2S-n_i} ~b_i + {\rm h.c.})/2$ and $S^z_i = (\sqrt{2S-n_i} ~b_i - {\rm h.c.})/(2i)$ ($n_i = b_i^\dagger b_i$), where $b_i, b_i^\dagger$ are bosonic operators; and then we move to momentum space for the bosonic operators,  $b_i = N^{-1/2} \sum_{\bm q} e^{i \bm q \cdot \bm r_i} b_{\bm q}$.
The XXZ Hamiltonian expressed in terms of HP bosons has a priori a highly non-linear form; yet the importance of non-linearities can be very different when looking at zero-momentum bosons versus finite-momentum ones. 
 By construction the $|{\rm CSS}_x\rangle$ state coincides with the vacuum of all HP bosons; and the dynamics initialized in this state has the major effect of depolarizing the collective spin, namely of letting $\langle J^x \rangle$ relax to zero, under proliferation of bosons. Ferromagnetic couplings for the $x$ and $y$ spin components imply that the lowest-energy bosons have zero momentum, so that one should expect a much faster proliferation of these bosons compared to finite-momentum ones, so that in practice for all times (except at the very start) $\langle b_0^\dagger b_0 \rangle \gg \langle b_{\bm q \neq 0}^\dagger b_{\bm q\neq 0} \rangle$. 
 Hence nonlinearities for the zero-momentum bosons should be handled with greatest care. As detailed in the Supplemental Material (SM) \cite{SM} (see also Ref.~\cite{Roscildeetal2023}), \emph{all} the terms in the bosonic Hamiltonian containing exclusively $b_0$ and $b_0^\dagger$ bosons can be resummed to reconstruct a planar-rotor (or OAT) model
 \begin{equation}
 {\cal H}_{\rm R} = E_{0,\rm R} + \frac{(K^z)^2}{2I}
 \label{e.rotor}
 \end{equation} 
 where $E_{0,\rm R}$ is the rotor ground-state energy,  and  ${\bm K}$ is an angular momentum operator of macroscopic length $NS$, associated with the zero-momentum bosons, namely $K^x = NS - b_0^\dagger b_0$, $K^y = (\sqrt{2NS-n_0}~ b_0 + {\rm h.c.})/2$ and  $K^z = (\sqrt{2NS-n_0}~ b_0 - {\rm h.c.})/(2i)$; and the moment of inertia of the rotor variable is given by $1/(2I) = J_{\bm q=0} (1-\Delta)/[2(N-1)]$ \cite{SM} where $J_{\bm q} = N^{-1} \sum_{ij} e^{i{\bm q} \cdot (\bm r_i - \bm r_j)} ~J_{ij} $ \footnote{The moment of inertia of the rotor can become negative when $\Delta>1$, signaling the fact that the low-energy physics of the system is no longer akin to that of a rotor in the $xy$ plane, because the system develops Ising-like ferromagnetism along $z$ instead.  Yet a negative moment of inertia is not an issue when considering the real-time dynamics.}.
  On the other hand, upon linearizing the Hamiltonian in terms of the finite-momentum bosons, one obtains 
\begin{equation}
{\cal H}_{\rm XXZ} = {\cal H}_{\rm R} + {\cal H}_{\rm SW} + {\cal O}(n_0 n_{\bm q \neq 0})
\label{e.RSW}
\end{equation}
where 
${\cal H}_{\rm SW} = \sum_{\bm q\neq 0} \left [ A_{\bm q} b^\dagger_{\bm q} b_{\bm q} + \frac{1}{2} B_{\bm q} \left (b_{\bm q} b_{-\bm q} + b^\dagger_{\bm q} b^\dagger_{-\bm q} \right ) \right ] $
is the quadratic SW Hamiltonian, with $A_{\bm q}  =   S \left [ J_0 - J_{\bm q}(1+\Delta)/2 \right] $ and $B_{\bm q}  =  - J_{\bm q} S (1-\Delta)/2$ \cite{Frerot2017PRB}. 

The central assumption of the RSW scheme is that the most important non-linearities in the system are all contained in ${\cal H}_{\rm R}$, while the further non-linear terms are negligible. This leads to the additive structure of Eq.~\eqref{e.RSW}, implying an effective separation between a non-linear rotor and linear SWs. 
Discarding the same kinds of terms for all the quantities of interest leads to a similarly additive structure: e.g. we obtain that $\langle J^x \rangle = \langle K^x \rangle - N_{\rm FM}$  where $N_{\rm FM} = \sum_{\bm q\neq 0} \langle b_{\bm q}^\dagger b_{\bm q} \rangle$ is the total number of finite-momentum (FM) bosons; ${\rm Var}(J^x) \approx {\rm Var}(K^x) - 2(NS-\langle K^x\rangle)N_{\rm FM} - N_{\rm FM}^2$ and ${\rm Var}(J^{\perp}) \approx {\rm Var}(K^\perp)$. 
When the ground state of the XXZ Hamiltonian breaks the $U(1)$ symmetry in the thermodynamic limit (\emph{e.g} for $|\Delta|<1$ in 2$d$), the low-lying spectrum is expected to feature  a so-called Anderson tower of states (ToS) \cite{Anderson1997,Tasaki2018JSP}, possessing the same structure as that of a OAT model, $E_{\rm ToS}(J^z) = E_0 + (J^z)^2/I_{\rm ToS}$. The fact that $\langle (J^z)^2 \rangle\approx \langle (K^z)^2 \rangle$ shows that the spectrum of the rotor Hamiltonian Eq.~\eqref{e.rotor} reconstructs explicitly the Anderson ToS, albeit with $I \neq I_{\rm ToS}$ in general. This discrepancy can be understood from the residual coupling between SW and rotor that we discard, and which can be thought of as renormalizing the bare moment of inertia of the rotor. In the following we shall redefine the rotor variable so as to take into account this renormalization, namely $I \to I_{\rm ToS}$. We detail in the SM \cite{SM} how to systematically reconstruct $I_{\rm ToS}$ for all the system sizes $N$ we considered. 

\emph{Dynamics of the dipolar XX model.} Within the RSW scheme, the quench dynamics of the system is then solved at a polynomial cost by evolving separately the rotor variable (with a $(2NS+1)$-dimensional Hilbert space) and the SW ones (whose dynamics can be solved for analytically \cite{Frerot2018PRL}). We then apply the RSW approach to the relevant case of the dipolar $S=1/2$ XX model, corresponding to $\Delta = 0$ and $J_{ij} = J |\bm r_i - \bm r_j|^{-3}$, in the case of a square lattice. This model is literally realized in Rydberg-atom arrays with resonant interactions \cite{BrowaeysL2020, Chenetal2022}, but many more platforms are described by XXZ dipolar models \cite{Hazzardetal2013,Chomazetal2022}.  Fig.~\ref{fig1} shows a systematic comparison of our RSW results with those of tVMC based on a pair-product Ansatz \cite{Thibaut2019PRB,Comparin2022PRA,Comparinetal2022b}. Fig.~\ref{fig1}(a) shows the depolarizing dynamics of the average collective spin $\langle J^x \rangle$, which admits an exact additive structure in terms of the zero-momentum and finite-momentum components. We observe that the population of zero-momentum bosons becomes very quickly much larger than that of all the finite-momentum ones taken together, validating the basic assumption of RSW theory. Subtracting the boson populations from the initial average spin $N/2$ gives a prediction for $\langle J^x \rangle$ in very good agreement with tVMC, especially at short times -- while at longer times the neglect of the coupling between rotor and SW leads to deviations from tVMC.  The short-time squeezing dynamics is then examined in Fig.~\ref{fig1}(b), showing that RSW and tVMC are in nearly perfect agreement. In fact, due to the very weak population of finite-momentum bosons, the squeezing parameter is almost fully accounted for by the rotor variable alone. The SW contribution enters uniquely via $N_{\rm FM}$ in $\langle J^x \rangle^2$; yet it provides the slight renormalization which fixes the discrepancy between the bare rotor results and the tVMC ones.

Fig.~\ref{fig1}(c) goes beyond the short-time squeezing dynamics, and looks instead at the evolution of ${\rm Var}(J^x)$, which attains first a plateau at $N^2/8$ in the OAT dynamics, followed by a sharp peak of height $N^2/4$ at time $\pi I$, which corresponds to the appearance of a GHZ state. The RSW results show that the SW contribution to the variance introduces a reduction in the plateau value as well as oscillations at the characteristic frequency of evolution of $N_{\rm FM}$. This behavior reflects closely that of the tVMC results, with a very clear correspondence in the fluctuating part, although the amplitude of the oscillations appears to be overestimated by the RSW results. Finally both the tVMC results and the RSW ones show the peak associated with the appearance of the GHZ-like state, with a height slightly reduced by the SW contribution, which quantitatively accounts for the deviation from the bare-rotor results.  The RSW scenario hence justifies the persistence of the formation of a GHZ-like state up to $N=144$ spins, as already reported by us in Ref.~\cite{Comparinetal2022b}. The formation of the GHZ-like state, along with that of $q$-headed cat states with $q>2$ at earlier times, can also be tracked by inspecting the half-system R\'enyi entanglement entropy, $S_2(N/2) = -\log {\rm Tr}(\rho^2_{N/2})$, where $\rho_{N/2}$ is the reduced state of a rectangle of $L\times L/2 = N/2$ spins. Within the RSW approach this entropy is strictly additive, and it is composed of a rotor contribution and a SW one \cite{Frerot2017PRB,Frerot2018PRL,Kurkjian2013PRA}. Fig.~\ref{fig1}(d) shows that the addition of the rotor and SW entropies accounts very closely for the tVMC results. In particular the rotor entropy exhibits a succession of dips occurring at times $t_q = 2\pi I/q$ corresponding to the formation of the $q$-headed cats \cite{Kurkjian2013PRA}, which is nicely reflected by the tVMC results as well, with superposed fluctuations coming from the SW contribution. It is worth noticing that the rotor entropy is ${\cal O}(\log N)$ (due to the polynomial Hilbert space dimensions for the rotor), while the maximum SW entropy is  ${\cal O}(N)$ (volume-law scaling); nonetheless the SW excitations are so dilute that the entropy is clearly dominated by the rotor contribution for the sizes ($N\sim 100$) considered here. 

\begin{figure}[ht!]
\includegraphics[width=\columnwidth]{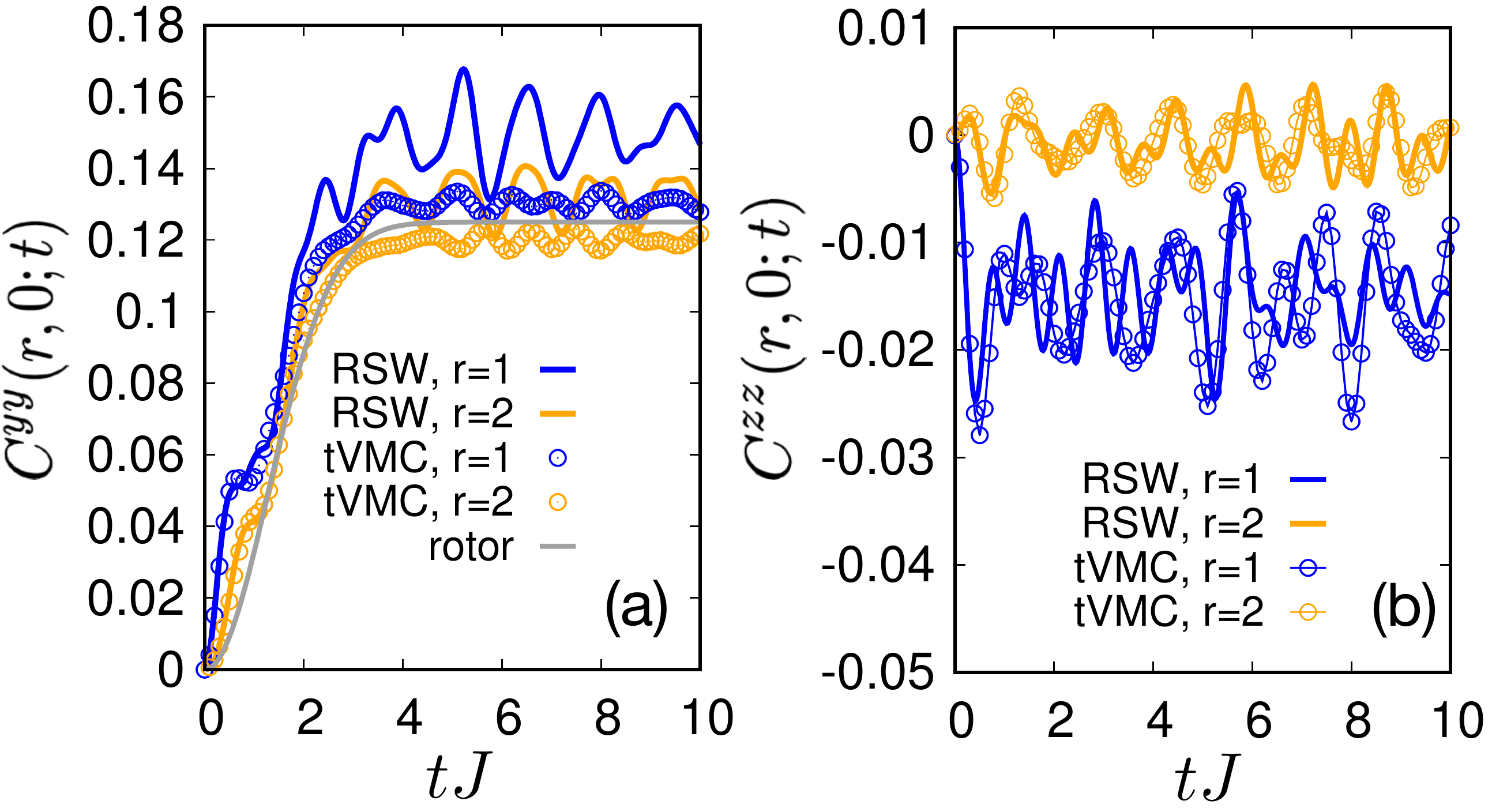}
\caption{\emph{Correlation dynamics for the 2$d$ dipolar XX model.} (a) $C^{yy}$ correlations at distance $r=1$ and 2; (b)  $C^{zz}$ correlations for the same distances.}
\label{fig2}
\end{figure}

A final element of comparison concerns the dynamics of correlations. Fig.~\ref{fig2} focuses in particular on the correlation functions $C^{zz}(\bm d) = \langle S_i^z S_{i+\bm d}^z \rangle$ and  $C^{yy}(\bm d) = \langle S_i^y S_{i+\bm d}^y \rangle$ for spin components perpendicular to the collective-spin orientation -- see also SM \cite{SM} for further extended data. The RSW approach reveals that these correlations possess a very distinct origin: {for the system size considered here ($N=100$)} the $C^{yy}$ correlations are dominated by the rotor contribution, $\langle (K^y)^2\rangle/N^2$, which is independent of the distance, while their weak spatial modulation comes from the SW contribution. On the other hand the $C^{zz}$ correlations are exclusively given by the SW contribution, because the rotor Hamiltonian commutes with the $S_i^z$ operators, and therefore correlations among them cannot develop under the rotor dynamics. This observation justifies why SW theory alone, neglecting any zero mode, can successfully describe the  $C^{zz}$ correlations \cite{Frerot2018PRL}. Fig.~\ref{fig2} shows that SW excitations add a spatial modulation and an oscillating behavior on top of the rotor contribution for the $C^{yy}$ correlations, while they fully account for the  $C^{zz}$ correlations; this justifies also why the latter correlations are roughly an order of magnitude smaller than the $C^{yy}$ ones. Our results suggest also that a Fourier analysis of the $C^{zz}$ correlations via quench spectroscopy \cite{Menu2018PRB,Villa2019PRA} allows one to reconstruct selectively the dispersion relation of the finite-momentum SW excitations; while the same analysis for the  $C^{yy}$ correlations would reveal as well the ToS excitations at zero momentum. {The fact that $C^{yy}$ correlations are dominated by the rotor dynamics is apparently in contradiction with the picture of correlation dynamics as being governed by propagating quasi-particles \cite{CalabreseC2006,Cheneauetal2012}; nonetheless the rotor dynamics is parametrically slower the larger the size $N$, so that correlation spreading keeps a causal structure, as further discussed in the SM \cite{SM}.}

\begin{figure}[ht!]
\includegraphics[width=0.7\columnwidth]{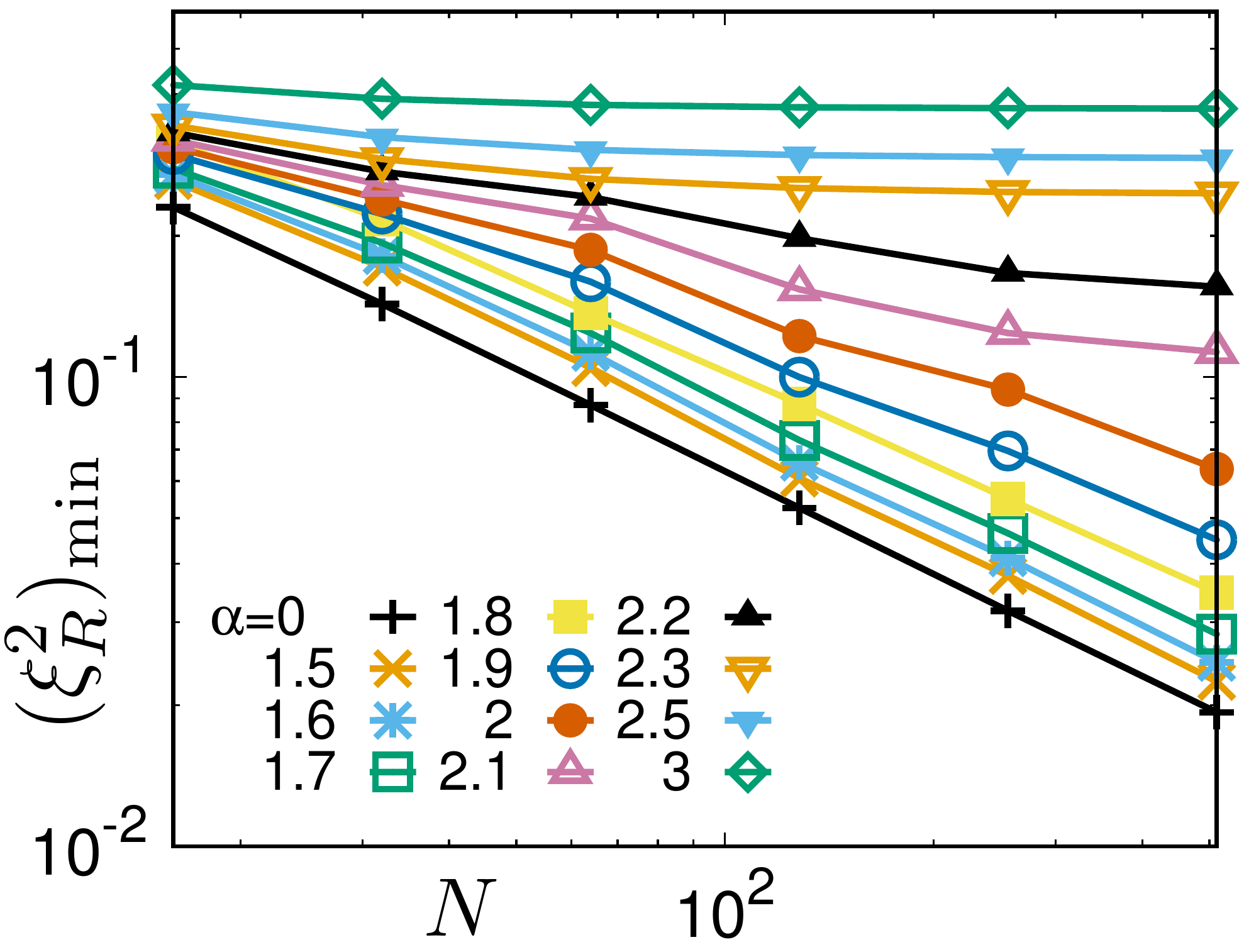}
\caption{\emph{Dynamical transition in the 1d XX model with power-law interactions.} Scaling of the minimum squeezing parameter for various values of the power $\alpha$ for the decay of interactions.}
\label{fig3}
\end{figure}

\emph{Dynamical transition in XX chains with power-law interactions.}  The above results have shown that the dynamics of the dipolar XX model in $d=2$ is dominated by the rotor contribution; this aspect justifies a posteriori the assumptions of RSW theory and explains its success for this specific example. Remarkably RSW theory can also signal the appearance of a dynamical transition from a OAT-like dynamics to a non-OAT one, when its assumptions fail at long times. We demonstrate this aspect in the case of the XX model with power-law interactions, $J_{ij} = J |r_i - r_j|^{-\alpha}$, cast on a $d=1$ lattice. As already shown in Refs.~\cite{Comparin2022PRA,Blocketal2023} this model exhibits OAT-like dynamics with scalable squeezing for $\alpha \lesssim 1.6$, and absence of scalable squeezing for larger values of $\alpha$. Fig.~\ref{fig3} shows the minimum value of the $\xi_R^2$ parameter achieved during the RSW dynamics \footnote{For these calculations, for the sake of simplicity we considered a rotor with bare moment of inertia.}. Two scaling behaviors are clearly exhibited: one compatible with OAT dynamics at small $\alpha$, and one compatible with non-scalable squeezing at large $\alpha$ -- with a seemingly smooth crossover between the two regimes occurring around $\alpha\approx 2$, showing that RSW theory overestimates the value of $\alpha$ at which scalable squeezing is lost. The non-scaling regime is due to the fact that the proliferation of SWs contribute to the average collective spin $\langle J^x \rangle$, in such a way that the depolarization happens faster than the onset of scaling for the minimum variance of the transverse spin components. The breakdown of OAT scaling in the squeezing signals that the RSW predictions cease to be quantitative at longer times. 
   
  \emph{Conclusions.} In this work we have demonstrated that the entangling dynamics of the infinite-range one-axis-twisting model can be reproduced with minor alterations by U(1)-symmetric spin models with power-law decaying interactions, thanks to an effective separation between zero-momentum degrees of freedom, possessing the spectrum of a planar-rotor variable; and finite-momentum ones, corresponding to spin-wave excitations. This effective separation of variable is always justified at short times, and it remains justified even for macroscopic (${\cal O}(N)$) times when the spin-wave excitations are weakly populated, allowing for the appearance of scalable spin squeezing, as well as of scalable cat-like states. The quantitative success of our approach in describing the dynamics of \emph{e.g.} dipolar spins in $2d$ suggests that this and similar systems, albeit not being properly integrable, undergo a rather peculiar dynamics at low energy. This dynamics is dominated by persistent spin-wave oscillations; and approximate recurrences at times ${\cal O}(N)$ -- as opposed to Poincar\'e times  ${\cal O}(\exp(N))$ -- which reflect the reduced Hilbert space of the rotor variable. While rotor and spin waves should eventually come to thermalize with each other thanks to their residual coupling, the time scales over which such thermalization occurs are currently unknown to us. Our findings suggest that effective rotor/spin-wave decoupling represents the mechanism by which a very large class of power-law interacting Hamiltonians implemented by quantum simulators -- including Rydberg atoms \cite{BrowaeysL2020}, magnetic atoms \cite{Chomazetal2022}, trapped ions \cite{Monroeetal2021}, superconducting circuits \cite{Song2019} -- can evade standard thermalization at low energy. And, by virtue of this mechanism, they can act as entangling resources producing scalable multipartite entanglement of interest for fundamental studies, as well as for potential metrological applications.

\begin{acknowledgements}
 \emph{Acknowledgements.} This work is supported by ANR (EELS project), QuantERA (MAQS project) and PEPR-Q (QubitAF project). Fruitful discussions with M. Block, G. Bornet, A. Browaeys, C. Cheng, G. Emperauger, T. Lahaye, B. Ye and N. Yao are gratefully acknowledged. All numerical simulations have been performed on the PSMN cluster at the ENS of Lyon.   
\end{acknowledgements}
 
\newpage 

\begin{center}
{\bf Supplemental Material: Entangling dynamics from effective rotor/spin-wave separation in U(1)-symmetric quantum spin models}
\end{center}

\section{Reconstructing the OAT model from zero-momentum bosons; relevant observables}

In this section we briefly illustrate how one goes from the XXZ Hamiltonian
 \begin{equation}
{\cal H}_{\rm XXZ} = - \sum_{i< j}  J_{ij} \left ( S_i^x S_j^x + S_i^y S_j^y + \Delta S_i^z S_j^z  \right) 
\end{equation}
to the effective Hamiltonian of Eq.~2 of the main text, comprising a one-axis-twisting (OAT) model for the zero-momentum degrees of freedom; and a linear spin-wave (SW) part for the finite-momentum ones. 

In principle, a separation between zero-momentum and finite-momentum degrees of freedom can be achieved by simple Fourier transformation of the spin operators
\begin{equation}
S_i^\alpha = \frac{1}{\sqrt{N}} \sum_{\bm q} e^{i\bm q\cdot \bm r_i} S_{\bm q}^\alpha ~~~~~~ (\alpha = x,y,z)
\end{equation}
which, when applied to translationally invariant interactions $J_{ij} = J_{\bm r_i - \bm r_j}$ leads to 
\begin{eqnarray}
{\cal H}_{\rm XXZ} & = &  -  \sum_{\bm q} \frac{J_{\bm q}}{2} \left ( S_{\bm q}^x  S_{-\bm q}^x +   S_{\bm q}^y  S_{-\bm q}^y + \Delta  S_{\bm q}^z  S_{-\bm q}^z \right ) \nonumber \\
&=& - \frac{J_0}{2N} \left [ (J^x)^2 + (J^x)^2 + \Delta (J^z)^2 \right ] \nonumber  \\
&& - \sum_{\bm q\neq 0} \frac{J_{\bm q}}{2} \left ( S_{\bm q}^x  S_{-\bm q}^x +   S_{\bm q}^y  S_{-\bm q}^y + \Delta  S_{\bm q}^z  S_{-\bm q}^z \right ) \nonumber
\label{e.naive_separation}
\end{eqnarray}
where we have introduced the Fourier transform of the couplings
\begin{equation}
J_{\bm q} = \frac{1}{N} \sum_{ij} e^{i {\bm q} \cdot ({\bm r}_i - {\bm r}_j)}  J_{ij}
\end{equation}
and used the fact that $S_0^\alpha = J^\alpha/\sqrt{N}$. Clearly Eq.~\eqref{e.naive_separation} gives to the XXZ Hamiltonian the form of a U(1) symmetric Hamiltonian for the collective spin ${\bm J}$; and further terms containing only the finite-momentum Fourier components of the spins. Yet this decomposition is not really useful, because the different Fourier components of the spins are non-commuting operators, $[S_0^\alpha, S_{\bm q}^\beta] \neq 0$ for $\alpha \neq \beta$. This is another way of saying that the collective spin ${\bm J}$ is not conserved by the XXZ Hamiltonian, unless the couplings are of infinite range, namely unless $J_{\bm q} = J_0~ \delta_{{\bm q}, 0}$.

\begin{figure*}[ht!]
\includegraphics[width=0.9\textwidth]{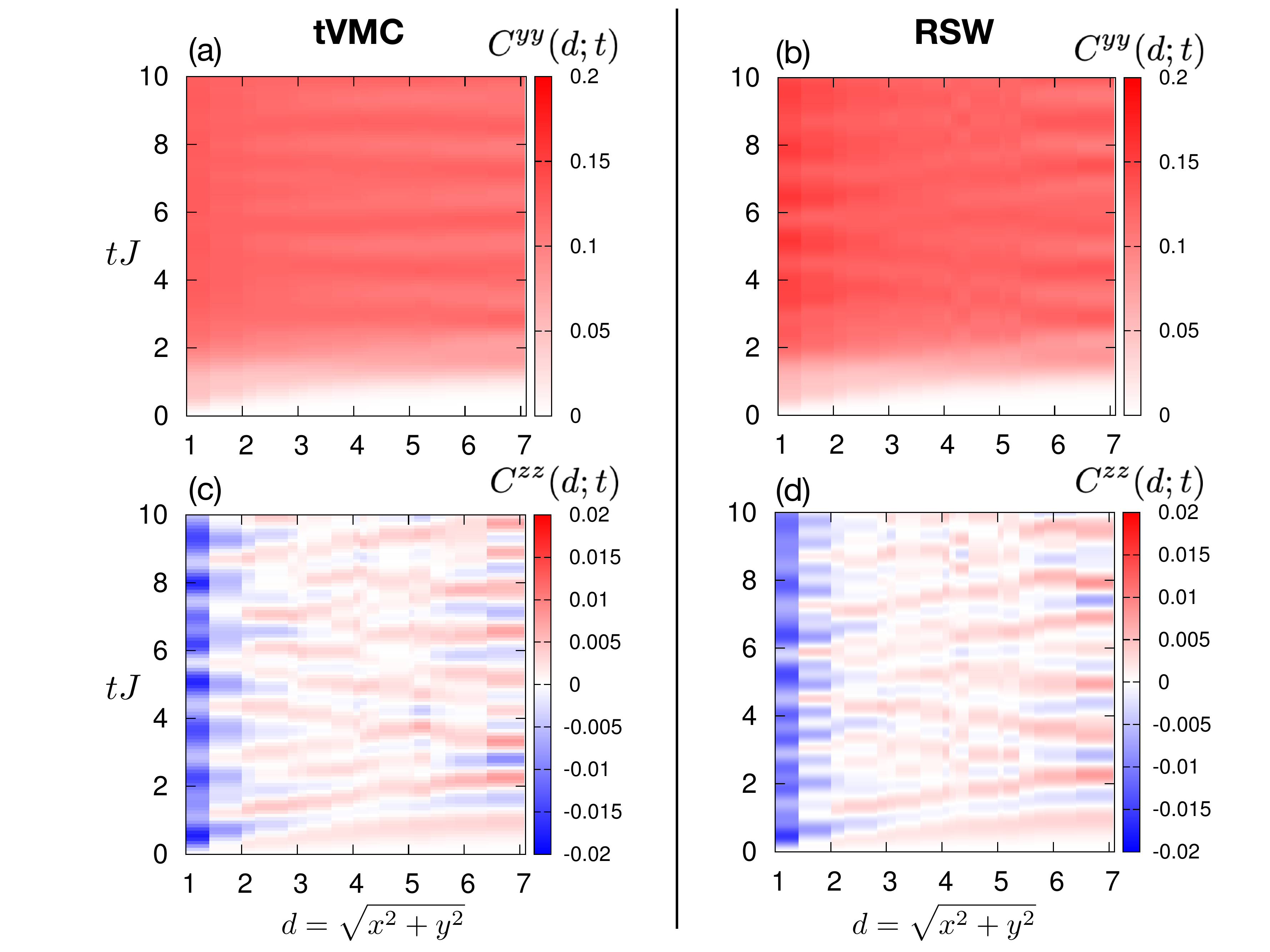}
\caption{\emph{Correlation dynamics for the 2$d$ dipolar XX model.} Evolution of the $C^{yy}$ (a-b) and $C^{zz}$ (c-d) correlations as a function of time for a system of $N=100$ spins, resulting from the tVMC calculation (left panels) and from the RSW approach (right panels).}
\label{fig4}
\end{figure*}

\begin{figure*}[ht!]
\includegraphics[width=0.9\textwidth]{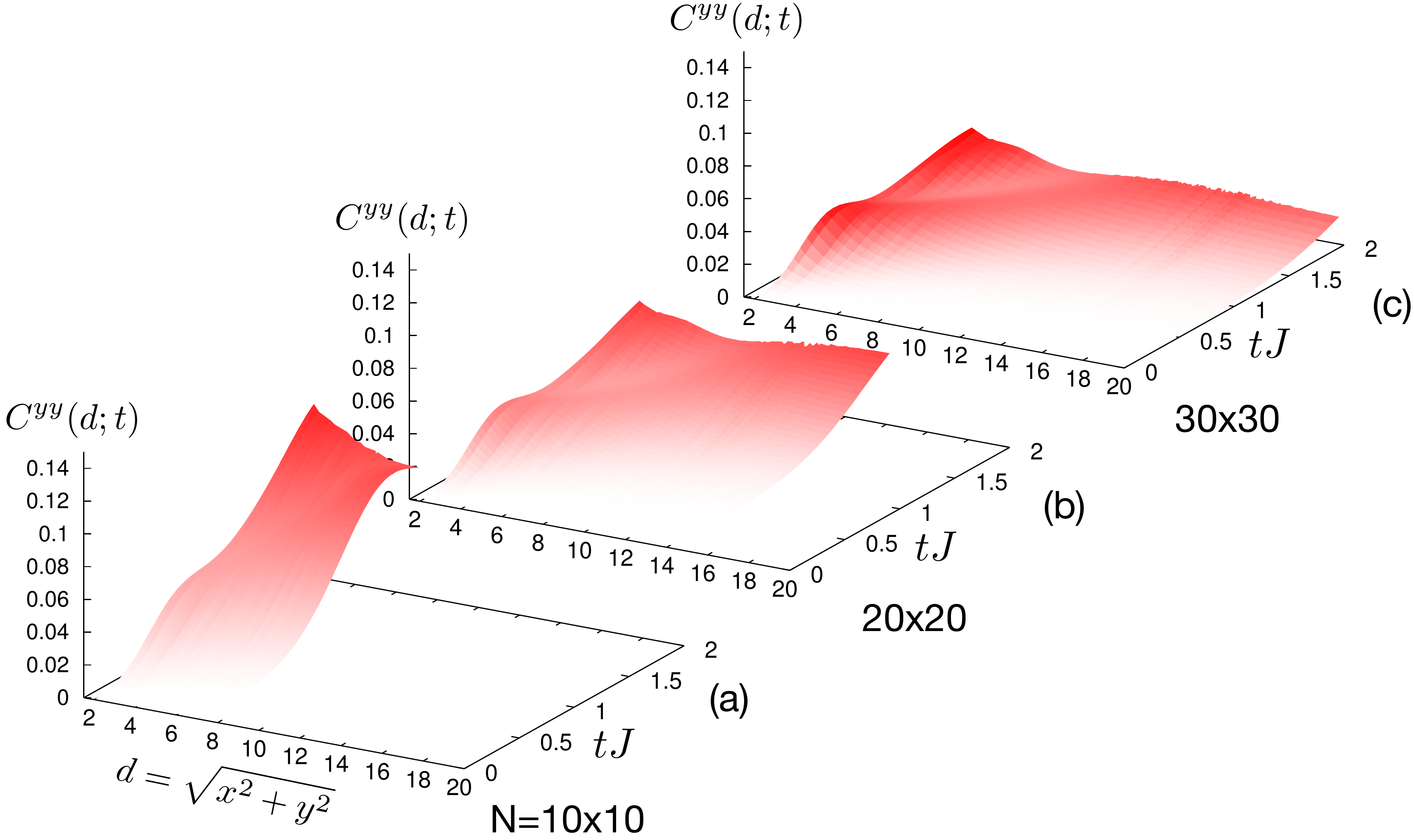}
\caption{\emph{Correlation dynamics for different system sizes.} The three panels show the evolution of $C^{yy}(x,y)$ for the 2d dipolar XX model from RSW theory, for growing systems sizes $N=10 \times 10$, $20\times 20$ and $30\times 30$.}
\label{fig5}
\end{figure*}

On the other hand, as already mentioned in the main text, a true separation of variables can be achieved by mapping spins onto Holstein-Primakoff (HP) bosons
\begin{eqnarray}
 S_i^x &=&  S - n_i \nonumber \\
 S^y_i &=& \frac{1}{2} (\sqrt{2S-n_i} ~b_i + {\rm h.c.}) \nonumber \\
 S^z_i &=& \frac{1}{2i} (\sqrt{2S-n_i} ~b_i - {\rm h.c.}) ~
 \label{e.HP}
 \end{eqnarray}
and by going from bosonic operators in real space to those in momentum space
\begin{equation}
b_i = \frac{1}{\sqrt{N}} \sum_{\bm q} e^{i \bm q \cdot \bm r_i} b_{\bm q}~.
\label{e.FT}
\end{equation}
Indeed bosonic operators associated with different momenta commute. And, for a translationally invariant Hamiltonian, it must be possible to isolate a part which contains \emph{exclusively} zero-momentum bosons $b_0, b_0^\dagger$; and a part which contains exclusively finite momentum ones in momentum conserving combinations. These two parts are then guaranteed to commute with each other. 
 For the choice of quantization axis in the HP transformation as in Eq.~\eqref{e.HP}, the XXZ Hamiltonian contains only even powers of HP bosons; and therefore the lowest-order momentum-conserving couplings between zero-momentum bosons and finite-momentum ones are of fourth order in the bosonic operators, and are of the kind $b_0^\dagger b_0 b_{\bm q} b_{-\bm q}  + {\rm h.c.}$, $b_0^\dagger b_0 b^\dagger_{\bm q} b_{\bm q}$, and $b_0^\dagger b^\dagger_0 b_{\bm q} b_{-\bm q}  + {\rm h.c.}$, namely they are all of order ${\cal O}(n_0 n_{\bm q\neq 0})$, as quoted in the main text. Neglecting these coupling terms (and those of higher order) leads to the approximate separation of variables at the basis of this work. 

In order to reconstruct the Hamiltonian contaning exclusively the zero-momentum bosons, it suffices to realize that, from Eq.~\eqref{e.FT}, $b_i  = b_0/\sqrt{N}$ + (FM terms); and from the HP transformation of Eq.~\eqref{e.HP}, one has that 
\begin{eqnarray}
 S_i^x &=&  S - \frac{b_0^\dagger b_0}{N} +  \text{(FM terms)}\nonumber \\
 S^y_i &=& \frac{1}{2N} \left (\sqrt{2NS-n_0} ~b_0+ {\rm h.c.} \right ) +  \text{(FM terms)} \nonumber \\
 S^z_i &=& \frac{1}{2iN} \left (\sqrt{2NS-n_0} ~b_0 - {\rm h.c.} \right ) +  \text{(FM terms)}  ~~~
 \label{e.HP2}
 \end{eqnarray}
where ``FM terms" indicates terms that contain at least one finite-momentum boson operator $b_{\bm q\neq 0}, b^\dagger_{\bm q\neq 0}$. 
Introducing the macroscopic spin ${\bm K}$ of length $NS$
\begin{eqnarray}
K^x & = & NS - b_0^\dagger b_0 \nonumber \\
K^y & = & \frac{1}{2} (\sqrt{2NS-n_0}~ b_0 + {\rm h.c.}) \nonumber \\
K^z & = & \frac{1}{2i} (\sqrt{2NS-n_0}~ b_0 - {\rm h.c.}) 
\end{eqnarray}
it is immediate to realize that $S_i^\alpha = K^\alpha/N$ + (FM terms), so that the zero-momentum part of the XXZ Hamiltonian takes the planar-rotor (or OAT) form 
\begin{eqnarray}
{\cal H}_{\rm R} & = &  - \frac{J_0}{2N} \left [(K^x)^2 + (K^y)^2 + \Delta( K^z)^2 \right ]  \nonumber \\
&=& - \frac{1}{2} J_0 S(NS+1) + \frac{J_0(1-\Delta)}{2N} (K^z)^2~.
\end{eqnarray}
The lowest-order terms at finite momentum reconstruct instead the well-known spin-wave Hamiltonian, which can be solved for by Bogolyubov diagonalization \cite{Frerot2017PRB, Frerot2018PRL}.  

Applying the same zero-momentum/finite-momentum decomposition to all observables, and neglecting terms of the same order as those neglected in the Hamiltonian, leads to the expressions for the observables as reported in the main text. A detailed discussion concerning correlation functions can be found in our companion work, Ref.~\cite{Roscildeetal2023}. As for the entanglement entropies, their expression within SW theory can be found in Refs.~\cite{Frerot2017PRB, Frerot2018PRL}; the entropy of the half-system associated with the rotor variable wass estimated  as the entropy of a half OAT model, whose dynamics can be reconstructed from the analytic knowledge of the reduced density matrix, as detailed in Ref.~\cite{Kurkjian2013PRA}.

\section{Renormalized moment of inertia of the rotor from the tower-of-states spectrum}

{The moment of inertia of the rotor obtained in the previous section reads 
\begin{equation}
\frac{1}{2I} = \frac{J_0(1-\Delta)}{2N}
\end{equation}
Yet in the main text we quote another expression, which has $N\to N-1$. This expression comes from considering that the rotor Hamiltonian can be viewed as the projection of the XXZ Hamiltonian on the sector of states which are symmetric under permutation of the lattice sites \cite{Comparin2022PRA}, and which is spanned by the Dicke states $|J_{\rm tot}=NS,M\rangle$, namely eigenstates of the total spin operators of ${\bm J}^2$ (with eigenvalue $J_{\rm tot}(J_{\rm tot}+1) = NS(NS+1)$) and $J^z$ (with eigenvalue $M$). Indeed the Hamiltonian containing exclusively the $b_0, b_0^\dagger$ operators is fully symmetric under permutation of the sites, as so are the $b_0, b_0^\dagger$ operators. Hence we can write
\begin{equation}
{\cal H}_{\rm R} = \sum_{M,M'} \langle J_{\rm tot}, M| {\cal H}_{\rm XXZ} | J_{\rm tot}, M' \rangle~ | J_{\rm tot}, M' \rangle \langle J_{\rm tot}, M | ~.
\end{equation}
It is rather straightforward to show that 
\begin{align}
 & \langle J_{\rm tot}, M| {\cal H}_{\rm XXZ} | J_{\rm tot}, M' \rangle =    \frac{J_0(1-\Delta)}{2(N-1)}  M^2  \delta_{M,M'}   \\
 & - \frac{J_0}{2} \left ( N S^2 + \frac{1-\Delta}{N-1} \langle J_{\rm tot}, M | \sum_i (S_i^z)^2  | J_{\rm tot}, M \rangle \right ) \delta_{M,M'}  \nonumber
\end{align}
For $S=1/2$ spins the second line of the above expression is a constant, so that the only dependence on $M$ is in the first line, where we read out the momentum of inertia of the planar rotor as declared in the main text.} 

As discussed in the main text, the bare moment of inertia of the rotor, $I= (N-1)/[J_0(1-\Delta)]$, does not coincide with that describing the spectrum of the Anderson tower of states (ToS), $E_{\rm ToS}(J^z) = E_0 + (J^z)^2/(2I_{\rm ToS})$, because the exact spectrum takes into account the residual coupling between the zero-momentum bosons and the finite-momentum ones. To account for this coupling in an effective way, we can can assume that it renormalizes statically the moment of inertia of the rotor, so as to lead to the replacement $I \to I_{\rm ToS}$. The value of $I_{\rm ToS}$ can be extracted from the exact diagonalization of the XXZ Hamiltonian, as we did in Ref.~\cite{Comparinetal2022b}; this is clearly an operation which is limited to small system sizes ($N=16$ in Ref.~\cite{Comparinetal2022b}). {In order to extract the renormalized moment of inertia for the rotor on much larger sizes $N$, inaccessible to exact diagonalization, we can postulate that the renormalization factor leading from $I$ to $I_{\rm Tos}$ 
\begin{equation}
\gamma = \frac{I_{\rm ToS}^{(N)}}{I^{(N)}} 
\end{equation}
is nearly size-independent. If we evaluate it for some reference size $N_{\rm ref}$ (=16 for our case), then we estimate the ToS moment of inertia for a larger size $N$ as 
\begin{equation}
I_{\rm ToS}^{(N)} \approx \gamma I^{(N)} = \frac{N-1}{N_{\rm ref}-1} \frac{J_0^{(N_{\rm ref})}}{J_0^{(N)}} I_{\rm ToS}^{(N_{\rm ref})}
\label{e.scaling}
\end{equation}
which is the scaling formula that was already obtained by us in Ref.~\cite{Comparinetal2022b} (albeit via  a different argument). 
Using the fact that $I_{\rm ToS}^{(N=16)} \approx 2.42 J^{-1}$ for the dipolar XX model on a square lattice \cite{Comparinetal2022b} (to be compared with the bare moment of inertia $I = (N-1)/[J_0(1-\Delta)] = 2.47 J^{-1}$), we reconstruct via the scaling formula of Eq.~\eqref{e.scaling} the effective moment of inertia of all the lattice sizes considered in this work. These values were compared in Ref.~\cite{Comparinetal2022b} with those extracted from the characteristic frequencies of the collective-spin dynamics calculated with tVMC, and turned out to be in very good agreement. Admittedly the renormalization effect in the 2d XX dipolar model is very weak ($\approx 2 \%$), but it can be more significant in other models we have examined.}

\section{Time evolution of correlations}
Fig.~\ref{fig4} shows extended data for the correlation dynamics of the dipolar , namely the entire spatio-temporal structure of the $C^{yy}(x,y)$ and $C^{zz}(x,y)$ correlations as a function of time and distance $d = \sqrt{x^2+y^2}$. The comparison between the tVMC results and the RSW predictions for a $N=100$ lattice shows a very good overall agreement. In particular the $C^{yy}$ correlations display a nearly flat structure with small ripples: this is the result of them being dominated by the rotor contribution, which by construction has no spatial structure, while the ripples come from the sub-dominant spin-wave contributions. On the other hand the $C^{zz}$ correlations stem exclusively from spin waves within the RSW picture, and indeed they have a strongly fluctuating and sign-changing structure. The latter reflects the fact that their integral, which gives ${\rm Var}(J^z)=N/4$, must be conserved in time: given that this value is already exhausted by the on-site term $\sum_i \langle (S_i^z)^2 \rangle$, the off-site correlations must necessarily s um up to zero, so that any positive correlation must be compensated by negative ones.  

{The fact that $C^{yy}$ correlations receive a rotor contribution -- as pointed out in the main text -- may cast doubts on their \emph{causal} structure. Indeed the rotor contribution is not associated with (linear) quasi-particles propagating with a well-defined group velocity, such as spin waves, but rather with non-linear excitations at zero momentum. Therefore its presence seems to undermine the well-established picture of light-cone spreading of correlations \cite{CalabreseC2006} coming from traveling quasi-particles. Yet a finite-size analysis of the RSW results -- given by Fig.~\ref{fig5} -- shows that in fact the causal structure of correlations is maintained. Indeed the rotor dynamics is slower the larger the size, as the moment of inertia of the rotor grows as $N$. The non-local rotor contribution to the $C^{yy}$ correlations, $\langle (K^y)^2\rangle/N$, reaches saturation at the value $1/8$  in a time $t_{\rm R}$ scaling as $\sqrt{N}$,  namely scaling as the linear size $L$ of the system in 2$d$ -- numerically we observe that $t_{\rm R} J \approx 0.3 L$. As a result, fixing a time frame of interest ($[0,2J^{-1}]$ in Fig.~\ref{fig5}), the rotor contribution is less and less significant, as seen by the suppression of correlations when increasing the system size. Fig.~\ref{fig5} shows moreover that the traveling spin-wave excitations propagate correlations systematically faster than the growth of the rotor contribution, reaching the edge of the system on a time scale which is much shorter than the one taken by the rotor to establish its contribution of $1/8$ to the correlations. If we assume the correlation front of spin-wave excitations to move super-ballistically, as suggested in the figure, namely as $d \sim t^{1/z}$ with $z = 1/2$ for the dipolar XX model \cite{Frerot2018PRL}, we have that it reaches the edge in a time $t_{\rm SW} \sim L^z$, parametrically shorter than $t_{\rm R}$. We would like to point out that correlation fronts in systems such as the one of interest may reveal their actual nature only for much larger sizes, and what appears to be a correlation front at short times turns out to be only a correlation maximum, enveloped by a much slower correlation front, as studied carefully in Ref.~\cite{Cevolanietal2019}. We postpone a detailed study of correlation dynamics on much bigger systems to future work.}

\bibliography{SWrotor.bib}

\begin{thebibliography}{43}
\expandafter\ifx\csname natexlab\endcsname\relax\def\natexlab#1{#1}\fi
\expandafter\ifx\csname bibnamefont\endcsname\relax
  \def\bibnamefont#1{#1}\fi
\expandafter\ifx\csname bibfnamefont\endcsname\relax
  \def\bibfnamefont#1{#1}\fi
\expandafter\ifx\csname citenamefont\endcsname\relax
  \def\citenamefont#1{#1}\fi
\expandafter\ifx\csname url\endcsname\relax
  \def\url#1{\texttt{#1}}\fi
\expandafter\ifx\csname urlprefix\endcsname\relax\def\urlprefix{URL }\fi
\providecommand{\bibinfo}[2]{#2}
\providecommand{\eprint}[2][]{\url{#2}}

\bibitem[{\citenamefont{Horodecki et~al.}(2009)\citenamefont{Horodecki,
  Horodecki, Horodecki, and Horodecki}}]{Horodeckietal2009}
\bibinfo{author}{\bibfnamefont{R.}~\bibnamefont{Horodecki}},
  \bibinfo{author}{\bibfnamefont{P.}~\bibnamefont{Horodecki}},
  \bibinfo{author}{\bibfnamefont{M.}~\bibnamefont{Horodecki}},
  \bibnamefont{and}
  \bibinfo{author}{\bibfnamefont{K.}~\bibnamefont{Horodecki}},
  \bibinfo{journal}{Rev. Mod. Phys.} \textbf{\bibinfo{volume}{81}},
  \bibinfo{pages}{865} (\bibinfo{year}{2009}),
  \urlprefix\url{http://link.aps.org/doi/10.1103/RevModPhys.81.865}.

\bibitem[{\citenamefont{G\"{u}hne and T{\'{o}}th}(2009)}]{Guehne_2009}
\bibinfo{author}{\bibfnamefont{O.}~\bibnamefont{G\"{u}hne}} \bibnamefont{and}
  \bibinfo{author}{\bibfnamefont{G.}~\bibnamefont{T{\'{o}}th}},
  \bibinfo{journal}{Physics Reports} \textbf{\bibinfo{volume}{474}},
  \bibinfo{pages}{1} (\bibinfo{year}{2009}),
  \urlprefix\url{https://doi.org/10.1016/j.physrep.2009.02.004}.

\bibitem[{\citenamefont{Pezz\`e et~al.}(2018)\citenamefont{Pezz\`e, Smerzi,
  Oberthaler, Schmied, and Treutlein}}]{Pezze2018RMP}
\bibinfo{author}{\bibfnamefont{L.}~\bibnamefont{Pezz\`e}},
  \bibinfo{author}{\bibfnamefont{A.}~\bibnamefont{Smerzi}},
  \bibinfo{author}{\bibfnamefont{M.~K.} \bibnamefont{Oberthaler}},
  \bibinfo{author}{\bibfnamefont{R.}~\bibnamefont{Schmied}}, \bibnamefont{and}
  \bibinfo{author}{\bibfnamefont{P.}~\bibnamefont{Treutlein}},
  \bibinfo{journal}{Rev. Mod. Phys.} \textbf{\bibinfo{volume}{90}},
  \bibinfo{pages}{035005} (\bibinfo{year}{2018}),
  \urlprefix\url{https://link.aps.org/doi/10.1103/RevModPhys.90.035005}.

\bibitem[{\citenamefont{Friis et~al.}(2019)\citenamefont{Friis, Vitagliano,
  Malik, and Huber}}]{Friisetal2019}
\bibinfo{author}{\bibfnamefont{N.}~\bibnamefont{Friis}},
  \bibinfo{author}{\bibfnamefont{G.}~\bibnamefont{Vitagliano}},
  \bibinfo{author}{\bibfnamefont{M.}~\bibnamefont{Malik}}, \bibnamefont{and}
  \bibinfo{author}{\bibfnamefont{M.}~\bibnamefont{Huber}},
  \bibinfo{journal}{Nature Reviews Physics} \textbf{\bibinfo{volume}{1}},
  \bibinfo{pages}{72} (\bibinfo{year}{2019}), ISSN \bibinfo{issn}{2522-5820},
  \urlprefix\url{https://doi.org/10.1038/s42254-018-0003-5}.

\bibitem[{\citenamefont{Georgescu et~al.}(2014)\citenamefont{Georgescu, Ashhab,
  and Nori}}]{Georgescuetal2014}
\bibinfo{author}{\bibfnamefont{I.~M.} \bibnamefont{Georgescu}},
  \bibinfo{author}{\bibfnamefont{S.}~\bibnamefont{Ashhab}}, \bibnamefont{and}
  \bibinfo{author}{\bibfnamefont{F.}~\bibnamefont{Nori}},
  \bibinfo{journal}{Rev. Mod. Phys.} \textbf{\bibinfo{volume}{86}},
  \bibinfo{pages}{153} (\bibinfo{year}{2014}),
  \urlprefix\url{https://link.aps.org/doi/10.1103/RevModPhys.86.153}.

\bibitem[{\citenamefont{Nielsen and Chuang}(2010)}]{IkeandMike}
\bibinfo{author}{\bibfnamefont{M.~A.} \bibnamefont{Nielsen}} \bibnamefont{and}
  \bibinfo{author}{\bibfnamefont{I.~L.} \bibnamefont{Chuang}},
  \emph{\bibinfo{title}{{Quantum Computation and Quantum Information}}}
  (\bibinfo{publisher}{Cambridge University Press}, \bibinfo{year}{2010}).

\bibitem[{\citenamefont{Gross and Bloch}(2017)}]{Gross2017}
\bibinfo{author}{\bibfnamefont{C.}~\bibnamefont{Gross}} \bibnamefont{and}
  \bibinfo{author}{\bibfnamefont{I.}~\bibnamefont{Bloch}},
  \bibinfo{journal}{Science} \textbf{\bibinfo{volume}{357}},
  \bibinfo{pages}{995} (\bibinfo{year}{2017}),
  \urlprefix\url{https://doi.org/10.1126/science.aal3837}.

\bibitem[{\citenamefont{Browaeys and Lahaye}(2020)}]{BrowaeysL2020}
\bibinfo{author}{\bibfnamefont{A.}~\bibnamefont{Browaeys}} \bibnamefont{and}
  \bibinfo{author}{\bibfnamefont{T.}~\bibnamefont{Lahaye}},
  \bibinfo{journal}{Nat. Phys.} \textbf{\bibinfo{volume}{16}},
  \bibinfo{pages}{132} (\bibinfo{year}{2020}),
  \urlprefix\url{https://doi.org/10.1038/s41567-019-0733-z}.

\bibitem[{\citenamefont{Sch{\"a}fer et~al.}(2020)\citenamefont{Sch{\"a}fer,
  Fukuhara, Sugawa, Takasu, and Takahashi}}]{Schaeferetal2020}
\bibinfo{author}{\bibfnamefont{F.}~\bibnamefont{Sch{\"a}fer}},
  \bibinfo{author}{\bibfnamefont{T.}~\bibnamefont{Fukuhara}},
  \bibinfo{author}{\bibfnamefont{S.}~\bibnamefont{Sugawa}},
  \bibinfo{author}{\bibfnamefont{Y.}~\bibnamefont{Takasu}}, \bibnamefont{and}
  \bibinfo{author}{\bibfnamefont{Y.}~\bibnamefont{Takahashi}},
  \bibinfo{journal}{Nature Reviews Physics} \textbf{\bibinfo{volume}{2}},
  \bibinfo{pages}{411} (\bibinfo{year}{2020}), ISSN \bibinfo{issn}{2522-5820},
  \urlprefix\url{https://doi.org/10.1038/s42254-020-0195-3}.

\bibitem[{\citenamefont{Kaufman and Ni}(2021)}]{Kaufman2021NP}
\bibinfo{author}{\bibfnamefont{A.}~\bibnamefont{Kaufman}} \bibnamefont{and}
  \bibinfo{author}{\bibfnamefont{K.-K.} \bibnamefont{Ni}},
  \bibinfo{journal}{Nat. Phys.}  (\bibinfo{year}{2021}), ISSN
  \bibinfo{issn}{1745-2473, 1745-2481},
  \urlprefix\url{https://www.nature.com/articles/s41567-021-01357-2}.

\bibitem[{\citenamefont{Monroe et~al.}(2021)\citenamefont{Monroe, Campbell,
  Duan, Gong, Gorshkov, Hess, Islam, Kim, Linke, Pagano
  et~al.}}]{Monroeetal2021}
\bibinfo{author}{\bibfnamefont{C.}~\bibnamefont{Monroe}},
  \bibinfo{author}{\bibfnamefont{W.~C.} \bibnamefont{Campbell}},
  \bibinfo{author}{\bibfnamefont{L.-M.} \bibnamefont{Duan}},
  \bibinfo{author}{\bibfnamefont{Z.-X.} \bibnamefont{Gong}},
  \bibinfo{author}{\bibfnamefont{A.~V.} \bibnamefont{Gorshkov}},
  \bibinfo{author}{\bibfnamefont{P.~W.} \bibnamefont{Hess}},
  \bibinfo{author}{\bibfnamefont{R.}~\bibnamefont{Islam}},
  \bibinfo{author}{\bibfnamefont{K.}~\bibnamefont{Kim}},
  \bibinfo{author}{\bibfnamefont{N.~M.} \bibnamefont{Linke}},
  \bibinfo{author}{\bibfnamefont{G.}~\bibnamefont{Pagano}},
  \bibnamefont{et~al.}, \bibinfo{journal}{Rev. Mod. Phys.}
  \textbf{\bibinfo{volume}{93}}, \bibinfo{pages}{025001}
  (\bibinfo{year}{2021}),
  \urlprefix\url{https://link.aps.org/doi/10.1103/RevModPhys.93.025001}.

\bibitem[{\citenamefont{Garc\'ia-Ripoll}(2022)}]{Juanjobook}
\bibinfo{author}{\bibfnamefont{J.}~\bibnamefont{Garc\'ia-Ripoll}},
  \emph{\bibinfo{title}{{Quantum Information and Quantum Optics with
  Superconducting Circuits}}} (\bibinfo{publisher}{Cambridge University Press},
  \bibinfo{year}{2022}).

\bibitem[{\citenamefont{Kaufman et~al.}(2016)\citenamefont{Kaufman, Tai, Lukin,
  Rispoli, Schittko, Preiss, and Greiner}}]{Kaufmanetal2016}
\bibinfo{author}{\bibfnamefont{A.~M.} \bibnamefont{Kaufman}},
  \bibinfo{author}{\bibfnamefont{M.~E.} \bibnamefont{Tai}},
  \bibinfo{author}{\bibfnamefont{A.}~\bibnamefont{Lukin}},
  \bibinfo{author}{\bibfnamefont{M.}~\bibnamefont{Rispoli}},
  \bibinfo{author}{\bibfnamefont{R.}~\bibnamefont{Schittko}},
  \bibinfo{author}{\bibfnamefont{P.~M.} \bibnamefont{Preiss}},
  \bibnamefont{and} \bibinfo{author}{\bibfnamefont{M.}~\bibnamefont{Greiner}},
  \bibinfo{journal}{Science} \textbf{\bibinfo{volume}{353}},
  \bibinfo{pages}{794} (\bibinfo{year}{2016}), ISSN \bibinfo{issn}{0036-8075},
  \urlprefix\url{http://science.sciencemag.org/content/353/6301/794}.

\bibitem[{\citenamefont{Brydges et~al.}(2019)\citenamefont{Brydges, Elben,
  Jurcevic, Vermersch, Maier, Lanyon, Zoller, Blatt, and Roos}}]{Brydges2019}
\bibinfo{author}{\bibfnamefont{T.}~\bibnamefont{Brydges}},
  \bibinfo{author}{\bibfnamefont{A.}~\bibnamefont{Elben}},
  \bibinfo{author}{\bibfnamefont{P.}~\bibnamefont{Jurcevic}},
  \bibinfo{author}{\bibfnamefont{B.}~\bibnamefont{Vermersch}},
  \bibinfo{author}{\bibfnamefont{C.}~\bibnamefont{Maier}},
  \bibinfo{author}{\bibfnamefont{B.~P.} \bibnamefont{Lanyon}},
  \bibinfo{author}{\bibfnamefont{P.}~\bibnamefont{Zoller}},
  \bibinfo{author}{\bibfnamefont{R.}~\bibnamefont{Blatt}}, \bibnamefont{and}
  \bibinfo{author}{\bibfnamefont{C.~F.} \bibnamefont{Roos}},
  \bibinfo{journal}{Science} \textbf{\bibinfo{volume}{364}},
  \bibinfo{pages}{260} (\bibinfo{year}{2019}),
  \urlprefix\url{https://doi.org/10.1126/science.aau4963}.

\bibitem[{\citenamefont{Kitagawa and Ueda}(1993)}]{Kitagawa1993PRA}
\bibinfo{author}{\bibfnamefont{M.}~\bibnamefont{Kitagawa}} \bibnamefont{and}
  \bibinfo{author}{\bibfnamefont{M.}~\bibnamefont{Ueda}},
  \bibinfo{journal}{Phys. Rev. A} \textbf{\bibinfo{volume}{47}},
  \bibinfo{pages}{5138} (\bibinfo{year}{1993}),
  \urlprefix\url{https://link.aps.org/doi/10.1103/PhysRevA.47.5138}.

\bibitem[{\citenamefont{Wineland et~al.}(1994)\citenamefont{Wineland,
  Bollinger, Itano, and Heinzen}}]{Wineland1994PRA}
\bibinfo{author}{\bibfnamefont{D.~J.} \bibnamefont{Wineland}},
  \bibinfo{author}{\bibfnamefont{J.~J.} \bibnamefont{Bollinger}},
  \bibinfo{author}{\bibfnamefont{W.~M.} \bibnamefont{Itano}}, \bibnamefont{and}
  \bibinfo{author}{\bibfnamefont{D.~J.} \bibnamefont{Heinzen}},
  \bibinfo{journal}{Phys. Rev. A} \textbf{\bibinfo{volume}{50}},
  \bibinfo{pages}{67} (\bibinfo{year}{1994}),
  \urlprefix\url{https://link.aps.org/doi/10.1103/PhysRevA.50.67}.

\bibitem[{\citenamefont{Agarwal et~al.}(1997)\citenamefont{Agarwal, Puri, and
  Singh}}]{Agarwal1997PRA}
\bibinfo{author}{\bibfnamefont{G.~S.} \bibnamefont{Agarwal}},
  \bibinfo{author}{\bibfnamefont{R.~R.} \bibnamefont{Puri}}, \bibnamefont{and}
  \bibinfo{author}{\bibfnamefont{R.~P.} \bibnamefont{Singh}},
  \bibinfo{journal}{Phys. Rev. A} \textbf{\bibinfo{volume}{56}},
  \bibinfo{pages}{2249} (\bibinfo{year}{1997}),
  \urlprefix\url{https://link.aps.org/doi/10.1103/PhysRevA.56.2249}.

\bibitem[{\citenamefont{Riedel et~al.}(2010)\citenamefont{Riedel, B\"{o}hi, Li,
  H\"{a}nsch, Sinatra, and Treutlein}}]{Riedel2010}
\bibinfo{author}{\bibfnamefont{M.~F.} \bibnamefont{Riedel}},
  \bibinfo{author}{\bibfnamefont{P.}~\bibnamefont{B\"{o}hi}},
  \bibinfo{author}{\bibfnamefont{Y.}~\bibnamefont{Li}},
  \bibinfo{author}{\bibfnamefont{T.~W.} \bibnamefont{H\"{a}nsch}},
  \bibinfo{author}{\bibfnamefont{A.}~\bibnamefont{Sinatra}}, \bibnamefont{and}
  \bibinfo{author}{\bibfnamefont{P.}~\bibnamefont{Treutlein}},
  \bibinfo{journal}{Nature} \textbf{\bibinfo{volume}{464}},
  \bibinfo{pages}{1170} (\bibinfo{year}{2010}),
  \urlprefix\url{https://doi.org/10.1038/nature08988}.

\bibitem[{\citenamefont{Gross}(2012)}]{Gross2012}
\bibinfo{author}{\bibfnamefont{C.}~\bibnamefont{Gross}},
  \bibinfo{journal}{Journal of Physics B: Atomic, Molecular and Optical
  Physics} \textbf{\bibinfo{volume}{45}}, \bibinfo{pages}{103001}
  (\bibinfo{year}{2012}),
  \urlprefix\url{https://dx.doi.org/10.1088/0953-4075/45/10/103001}.

\bibitem[{\citenamefont{Norcia et~al.}(2018)\citenamefont{Norcia, Lewis-Swan,
  Cline, Zhu, Rey, and Thompson}}]{Norciaetal2018}
\bibinfo{author}{\bibfnamefont{M.~A.} \bibnamefont{Norcia}},
  \bibinfo{author}{\bibfnamefont{R.~J.} \bibnamefont{Lewis-Swan}},
  \bibinfo{author}{\bibfnamefont{J.~R.~K.} \bibnamefont{Cline}},
  \bibinfo{author}{\bibfnamefont{B.}~\bibnamefont{Zhu}},
  \bibinfo{author}{\bibfnamefont{A.~M.} \bibnamefont{Rey}}, \bibnamefont{and}
  \bibinfo{author}{\bibfnamefont{J.~K.} \bibnamefont{Thompson}},
  \bibinfo{journal}{Science} \textbf{\bibinfo{volume}{361}},
  \bibinfo{pages}{259} (\bibinfo{year}{2018}),
  \eprint{https://www.science.org/doi/pdf/10.1126/science.aar3102},
  \urlprefix\url{https://www.science.org/doi/abs/10.1126/science.aar3102}.

\bibitem[{\citenamefont{Braverman et~al.}(2019)\citenamefont{Braverman,
  Kawasaki, Pedrozo-Pe\~nafiel, Colombo, Shu, Li, Mendez, Yamoah, Salvi,
  Akamatsu et~al.}}]{Bravermanetal2019}
\bibinfo{author}{\bibfnamefont{B.}~\bibnamefont{Braverman}},
  \bibinfo{author}{\bibfnamefont{A.}~\bibnamefont{Kawasaki}},
  \bibinfo{author}{\bibfnamefont{E.}~\bibnamefont{Pedrozo-Pe\~nafiel}},
  \bibinfo{author}{\bibfnamefont{S.}~\bibnamefont{Colombo}},
  \bibinfo{author}{\bibfnamefont{C.}~\bibnamefont{Shu}},
  \bibinfo{author}{\bibfnamefont{Z.}~\bibnamefont{Li}},
  \bibinfo{author}{\bibfnamefont{E.}~\bibnamefont{Mendez}},
  \bibinfo{author}{\bibfnamefont{M.}~\bibnamefont{Yamoah}},
  \bibinfo{author}{\bibfnamefont{L.}~\bibnamefont{Salvi}},
  \bibinfo{author}{\bibfnamefont{D.}~\bibnamefont{Akamatsu}},
  \bibnamefont{et~al.}, \bibinfo{journal}{Phys. Rev. Lett.}
  \textbf{\bibinfo{volume}{122}}, \bibinfo{pages}{223203}
  (\bibinfo{year}{2019}),
  \urlprefix\url{https://link.aps.org/doi/10.1103/PhysRevLett.122.223203}.

\bibitem[{\citenamefont{Song et~al.}(2019)\citenamefont{Song, Xu, Li, Zhang,
  Zhang, Liu, Guo, Wang, Ren, Hao et~al.}}]{Song2019}
\bibinfo{author}{\bibfnamefont{C.}~\bibnamefont{Song}},
  \bibinfo{author}{\bibfnamefont{K.}~\bibnamefont{Xu}},
  \bibinfo{author}{\bibfnamefont{H.}~\bibnamefont{Li}},
  \bibinfo{author}{\bibfnamefont{Y.-R.} \bibnamefont{Zhang}},
  \bibinfo{author}{\bibfnamefont{X.}~\bibnamefont{Zhang}},
  \bibinfo{author}{\bibfnamefont{W.}~\bibnamefont{Liu}},
  \bibinfo{author}{\bibfnamefont{Q.}~\bibnamefont{Guo}},
  \bibinfo{author}{\bibfnamefont{Z.}~\bibnamefont{Wang}},
  \bibinfo{author}{\bibfnamefont{W.}~\bibnamefont{Ren}},
  \bibinfo{author}{\bibfnamefont{J.}~\bibnamefont{Hao}}, \bibnamefont{et~al.},
  \bibinfo{journal}{Science} \textbf{\bibinfo{volume}{365}},
  \bibinfo{pages}{574} (\bibinfo{year}{2019}),
  \urlprefix\url{https://doi.org/10.1126/science.aay0600}.

\bibitem[{\citenamefont{Foss-Feig et~al.}(2016)\citenamefont{Foss-Feig, Gong,
  Gorshkov, and Clark}}]{FossFeig2016Arxiv}
\bibinfo{author}{\bibfnamefont{M.}~\bibnamefont{Foss-Feig}},
  \bibinfo{author}{\bibfnamefont{Z.-X.} \bibnamefont{Gong}},
  \bibinfo{author}{\bibfnamefont{A.~V.} \bibnamefont{Gorshkov}},
  \bibnamefont{and} \bibinfo{author}{\bibfnamefont{C.~W.} \bibnamefont{Clark}},
  \emph{\bibinfo{title}{{Entanglement and spin-squeezing without infinite-range
  interactions}}} (\bibinfo{year}{2016}), \eprint{1612.07805},
  \urlprefix\url{https://arxiv.org/abs/1612.07805}.

\bibitem[{\citenamefont{Perlin et~al.}(2020)\citenamefont{Perlin, Qu, and
  Rey}}]{Perlin2020PRL}
\bibinfo{author}{\bibfnamefont{M.~A.} \bibnamefont{Perlin}},
  \bibinfo{author}{\bibfnamefont{C.}~\bibnamefont{Qu}}, \bibnamefont{and}
  \bibinfo{author}{\bibfnamefont{A.~M.} \bibnamefont{Rey}},
  \bibinfo{journal}{Phys. Rev. Lett.} \textbf{\bibinfo{volume}{125}},
  \bibinfo{pages}{223401} (\bibinfo{year}{2020}),
  \urlprefix\url{https://link.aps.org/doi/10.1103/PhysRevLett.125.223401}.

\bibitem[{\citenamefont{Comparin
  et~al.}(2022{\natexlab{a}})\citenamefont{Comparin, Mezzacapo, and
  Roscilde}}]{Comparin2022PRA}
\bibinfo{author}{\bibfnamefont{T.}~\bibnamefont{Comparin}},
  \bibinfo{author}{\bibfnamefont{F.}~\bibnamefont{Mezzacapo}},
  \bibnamefont{and} \bibinfo{author}{\bibfnamefont{T.}~\bibnamefont{Roscilde}},
  \bibinfo{journal}{Phys. Rev. A} \textbf{\bibinfo{volume}{105}},
  \bibinfo{pages}{022625} (\bibinfo{year}{2022}{\natexlab{a}}),
  \urlprefix\url{https://link.aps.org/doi/10.1103/PhysRevA.105.022625}.

\bibitem[{\citenamefont{Comparin
  et~al.}(2022{\natexlab{b}})\citenamefont{Comparin, Mezzacapo, and
  Roscilde}}]{Comparinetal2022b}
\bibinfo{author}{\bibfnamefont{T.}~\bibnamefont{Comparin}},
  \bibinfo{author}{\bibfnamefont{F.}~\bibnamefont{Mezzacapo}},
  \bibnamefont{and} \bibinfo{author}{\bibfnamefont{T.}~\bibnamefont{Roscilde}},
  \bibinfo{journal}{Phys. Rev. Lett.} \textbf{\bibinfo{volume}{129}},
  \bibinfo{pages}{150503} (\bibinfo{year}{2022}{\natexlab{b}}),
  \urlprefix\url{https://link.aps.org/doi/10.1103/PhysRevLett.129.150503}.

\bibitem[{\citenamefont{Block et~al.}(2023)\citenamefont{Block, Ye, Roberts,
  Chern, Wu, Wang, Pollet, Davis, Halperin, and Yao}}]{Blocketal2023}
\bibinfo{author}{\bibfnamefont{M.}~\bibnamefont{Block}},
  \bibinfo{author}{\bibfnamefont{B.}~\bibnamefont{Ye}},
  \bibinfo{author}{\bibfnamefont{B.}~\bibnamefont{Roberts}},
  \bibinfo{author}{\bibfnamefont{S.}~\bibnamefont{Chern}},
  \bibinfo{author}{\bibfnamefont{W.}~\bibnamefont{Wu}},
  \bibinfo{author}{\bibfnamefont{Z.}~\bibnamefont{Wang}},
  \bibinfo{author}{\bibfnamefont{L.}~\bibnamefont{Pollet}},
  \bibinfo{author}{\bibfnamefont{E.~J.} \bibnamefont{Davis}},
  \bibinfo{author}{\bibfnamefont{B.~I.} \bibnamefont{Halperin}},
  \bibnamefont{and} \bibinfo{author}{\bibfnamefont{N.~Y.} \bibnamefont{Yao}},
  \emph{\bibinfo{title}{A universal theory of spin squeezing}}
  (\bibinfo{year}{2023}), \urlprefix\url{https://arxiv.org/abs/2301.09636}.

\bibitem[{SM()}]{SM}
\bibinfo{note}{See {S}upplemental {M}aterial ({SM}) for details about: 1) ...}

\bibitem[{\citenamefont{Roscilde et~al.}(2023)\citenamefont{Roscilde, Comparin,
  and Mezzacapo}}]{Roscildeetal2023}
\bibinfo{author}{\bibfnamefont{T.}~\bibnamefont{Roscilde}},
  \bibinfo{author}{\bibfnamefont{T.}~\bibnamefont{Comparin}}, \bibnamefont{and}
  \bibinfo{author}{\bibfnamefont{F.}~\bibnamefont{Mezzacapo}},
  \emph{\bibinfo{title}{{Rotor/spin-wave theory for quantum spin models with
  U(1) symmetry -- companion paper}}} (\bibinfo{year}{2023}).

\bibitem[{\citenamefont{Fr\'erot et~al.}(2017)\citenamefont{Fr\'erot, Naldesi,
  and Roscilde}}]{Frerot2017PRB}
\bibinfo{author}{\bibfnamefont{I.}~\bibnamefont{Fr\'erot}},
  \bibinfo{author}{\bibfnamefont{P.}~\bibnamefont{Naldesi}}, \bibnamefont{and}
  \bibinfo{author}{\bibfnamefont{T.}~\bibnamefont{Roscilde}},
  \bibinfo{journal}{Phys. Rev. B} \textbf{\bibinfo{volume}{95}},
  \bibinfo{pages}{245111} (\bibinfo{year}{2017}),
  \urlprefix\url{https://link.aps.org/doi/10.1103/PhysRevB.95.245111}.

\bibitem[{\citenamefont{Anderson}(1997)}]{Anderson1997}
\bibinfo{author}{\bibfnamefont{P.~W.} \bibnamefont{Anderson}},
  \emph{\bibinfo{title}{{Basic Notions of Condensed Matter Physics}}}
  (\bibinfo{publisher}{Taylor \& Francis}, \bibinfo{address}{Boca Raton (FL)},
  \bibinfo{year}{1997}).

\bibitem[{\citenamefont{Tasaki}(2018)}]{Tasaki2018JSP}
\bibinfo{author}{\bibfnamefont{H.}~\bibnamefont{Tasaki}}, \bibinfo{journal}{J.
  Stat. Phys.} \textbf{\bibinfo{volume}{174}}, \bibinfo{pages}{735}
  (\bibinfo{year}{2018}),
  \urlprefix\url{https://doi.org/10.1007/s10955-018-2193-8}.

\bibitem[{\citenamefont{Fr\'erot et~al.}(2018)\citenamefont{Fr\'erot, Naldesi,
  and Roscilde}}]{Frerot2018PRL}
\bibinfo{author}{\bibfnamefont{I.}~\bibnamefont{Fr\'erot}},
  \bibinfo{author}{\bibfnamefont{P.}~\bibnamefont{Naldesi}}, \bibnamefont{and}
  \bibinfo{author}{\bibfnamefont{T.}~\bibnamefont{Roscilde}},
  \bibinfo{journal}{Phys. Rev. Lett.} \textbf{\bibinfo{volume}{120}},
  \bibinfo{pages}{050401} (\bibinfo{year}{2018}),
  \urlprefix\url{https://link.aps.org/doi/10.1103/PhysRevLett.120.050401}.

\bibitem[{\citenamefont{Chen et~al.}(2022)\citenamefont{Chen, Bornet, Bintz,
  Emperauger, Leclerc, Liu, Scholl, Barredo, Hauschild, Chatterjee
  et~al.}}]{Chenetal2022}
\bibinfo{author}{\bibfnamefont{C.}~\bibnamefont{Chen}},
  \bibinfo{author}{\bibfnamefont{G.}~\bibnamefont{Bornet}},
  \bibinfo{author}{\bibfnamefont{M.}~\bibnamefont{Bintz}},
  \bibinfo{author}{\bibfnamefont{G.}~\bibnamefont{Emperauger}},
  \bibinfo{author}{\bibfnamefont{L.}~\bibnamefont{Leclerc}},
  \bibinfo{author}{\bibfnamefont{V.~S.} \bibnamefont{Liu}},
  \bibinfo{author}{\bibfnamefont{P.}~\bibnamefont{Scholl}},
  \bibinfo{author}{\bibfnamefont{D.}~\bibnamefont{Barredo}},
  \bibinfo{author}{\bibfnamefont{J.}~\bibnamefont{Hauschild}},
  \bibinfo{author}{\bibfnamefont{S.}~\bibnamefont{Chatterjee}},
  \bibnamefont{et~al.}, \emph{\bibinfo{title}{Continuous symmetry breaking in a
  two-dimensional rydberg array}} (\bibinfo{year}{2022}),
  \urlprefix\url{https://arxiv.org/abs/2207.12930}.

\bibitem[{\citenamefont{Hazzard et~al.}(2013)\citenamefont{Hazzard, Manmana,
  Foss-Feig, and Rey}}]{Hazzardetal2013}
\bibinfo{author}{\bibfnamefont{K.~R.~A.} \bibnamefont{Hazzard}},
  \bibinfo{author}{\bibfnamefont{S.~R.} \bibnamefont{Manmana}},
  \bibinfo{author}{\bibfnamefont{M.}~\bibnamefont{Foss-Feig}},
  \bibnamefont{and} \bibinfo{author}{\bibfnamefont{A.~M.} \bibnamefont{Rey}},
  \bibinfo{journal}{Phys. Rev. Lett.} \textbf{\bibinfo{volume}{110}},
  \bibinfo{pages}{075301} (\bibinfo{year}{2013}),
  \urlprefix\url{https://link.aps.org/doi/10.1103/PhysRevLett.110.075301}.

\bibitem[{\citenamefont{Chomaz et~al.}(2022)\citenamefont{Chomaz,
  Ferrier-Barbut, Ferlaino, Laburthe-Tolra, Lev, and Pfau}}]{Chomazetal2022}
\bibinfo{author}{\bibfnamefont{L.}~\bibnamefont{Chomaz}},
  \bibinfo{author}{\bibfnamefont{I.}~\bibnamefont{Ferrier-Barbut}},
  \bibinfo{author}{\bibfnamefont{F.}~\bibnamefont{Ferlaino}},
  \bibinfo{author}{\bibfnamefont{B.}~\bibnamefont{Laburthe-Tolra}},
  \bibinfo{author}{\bibfnamefont{B.~L.} \bibnamefont{Lev}}, \bibnamefont{and}
  \bibinfo{author}{\bibfnamefont{T.}~\bibnamefont{Pfau}},
  \emph{\bibinfo{title}{Dipolar physics: A review of experiments with magnetic
  quantum gases}} (\bibinfo{year}{2022}),
  \urlprefix\url{https://arxiv.org/abs/2201.02672}.

\bibitem[{\citenamefont{Thibaut et~al.}(2019)\citenamefont{Thibaut, Roscilde,
  and Mezzacapo}}]{Thibaut2019PRB}
\bibinfo{author}{\bibfnamefont{J.}~\bibnamefont{Thibaut}},
  \bibinfo{author}{\bibfnamefont{T.}~\bibnamefont{Roscilde}}, \bibnamefont{and}
  \bibinfo{author}{\bibfnamefont{F.}~\bibnamefont{Mezzacapo}},
  \bibinfo{journal}{Phys. Rev. B} \textbf{\bibinfo{volume}{100}},
  \bibinfo{pages}{155148} (\bibinfo{year}{2019}),
  \urlprefix\url{https://link.aps.org/doi/10.1103/PhysRevB.100.155148}.

\bibitem[{\citenamefont{Kurkjian et~al.}(2013)\citenamefont{Kurkjian,
  Paw\l{}owski, Sinatra, and Treutlein}}]{Kurkjian2013PRA}
\bibinfo{author}{\bibfnamefont{H.}~\bibnamefont{Kurkjian}},
  \bibinfo{author}{\bibfnamefont{K.}~\bibnamefont{Paw\l{}owski}},
  \bibinfo{author}{\bibfnamefont{A.}~\bibnamefont{Sinatra}}, \bibnamefont{and}
  \bibinfo{author}{\bibfnamefont{P.}~\bibnamefont{Treutlein}},
  \bibinfo{journal}{Phys. Rev. A} \textbf{\bibinfo{volume}{88}},
  \bibinfo{pages}{043605} (\bibinfo{year}{2013}),
  \urlprefix\url{https://link.aps.org/doi/10.1103/PhysRevA.88.043605}.

\bibitem[{\citenamefont{Menu and Roscilde}(2018)}]{Menu2018PRB}
\bibinfo{author}{\bibfnamefont{R.}~\bibnamefont{Menu}} \bibnamefont{and}
  \bibinfo{author}{\bibfnamefont{T.}~\bibnamefont{Roscilde}},
  \bibinfo{journal}{Phys. Rev. B} \textbf{\bibinfo{volume}{98}},
  \bibinfo{pages}{205145} (\bibinfo{year}{2018}),
  \urlprefix\url{https://link.aps.org/doi/10.1103/PhysRevB.98.205145}.

\bibitem[{\citenamefont{Villa et~al.}(2019)\citenamefont{Villa, Despres, and
  Sanchez-Palencia}}]{Villa2019PRA}
\bibinfo{author}{\bibfnamefont{L.}~\bibnamefont{Villa}},
  \bibinfo{author}{\bibfnamefont{J.}~\bibnamefont{Despres}}, \bibnamefont{and}
  \bibinfo{author}{\bibfnamefont{L.}~\bibnamefont{Sanchez-Palencia}},
  \bibinfo{journal}{Phys. Rev. A} \textbf{\bibinfo{volume}{100}},
  \bibinfo{pages}{063632} (\bibinfo{year}{2019}),
  \urlprefix\url{https://link.aps.org/doi/10.1103/PhysRevA.100.063632}.

\bibitem[{\citenamefont{Calabrese and Cardy}(2006)}]{CalabreseC2006}
\bibinfo{author}{\bibfnamefont{P.}~\bibnamefont{Calabrese}} \bibnamefont{and}
  \bibinfo{author}{\bibfnamefont{J.}~\bibnamefont{Cardy}},
  \bibinfo{journal}{Phys. Rev. Lett.} \textbf{\bibinfo{volume}{96}},
  \bibinfo{pages}{136801} (\bibinfo{year}{2006}),
  \urlprefix\url{https://link.aps.org/doi/10.1103/PhysRevLett.96.136801}.

\bibitem[{\citenamefont{Cheneau et~al.}(2012)\citenamefont{Cheneau, Barmettler,
  Poletti, Endres, Schau{\ss}, Fukuhara, Gross, Bloch, Kollath, and
  Kuhr}}]{Cheneauetal2012}
\bibinfo{author}{\bibfnamefont{M.}~\bibnamefont{Cheneau}},
  \bibinfo{author}{\bibfnamefont{P.}~\bibnamefont{Barmettler}},
  \bibinfo{author}{\bibfnamefont{D.}~\bibnamefont{Poletti}},
  \bibinfo{author}{\bibfnamefont{M.}~\bibnamefont{Endres}},
  \bibinfo{author}{\bibfnamefont{P.}~\bibnamefont{Schau{\ss}}},
  \bibinfo{author}{\bibfnamefont{T.}~\bibnamefont{Fukuhara}},
  \bibinfo{author}{\bibfnamefont{C.}~\bibnamefont{Gross}},
  \bibinfo{author}{\bibfnamefont{I.}~\bibnamefont{Bloch}},
  \bibinfo{author}{\bibfnamefont{C.}~\bibnamefont{Kollath}}, \bibnamefont{and}
  \bibinfo{author}{\bibfnamefont{S.}~\bibnamefont{Kuhr}},
  \bibinfo{journal}{Nature} \textbf{\bibinfo{volume}{481}},
  \bibinfo{pages}{484} (\bibinfo{year}{2012}), ISSN \bibinfo{issn}{1476-4687},
  \urlprefix\url{https://doi.org/10.1038/nature10748}.

\bibitem[{\citenamefont{Cevolani et~al.}(2018)\citenamefont{Cevolani, Despres,
  Carleo, Tagliacozzo, and Sanchez-Palencia}}]{Cevolanietal2019}
\bibinfo{author}{\bibfnamefont{L.}~\bibnamefont{Cevolani}},
  \bibinfo{author}{\bibfnamefont{J.}~\bibnamefont{Despres}},
  \bibinfo{author}{\bibfnamefont{G.}~\bibnamefont{Carleo}},
  \bibinfo{author}{\bibfnamefont{L.}~\bibnamefont{Tagliacozzo}},
  \bibnamefont{and}
  \bibinfo{author}{\bibfnamefont{L.}~\bibnamefont{Sanchez-Palencia}},
  \bibinfo{journal}{Phys. Rev. B} \textbf{\bibinfo{volume}{98}},
  \bibinfo{pages}{024302} (\bibinfo{year}{2018}),
  \urlprefix\url{https://link.aps.org/doi/10.1103/PhysRevB.98.024302}.

\end{thebibliography}

\end{document}